\documentclass{article}

\usepackage[square,sort,comma,numbers]{natbib}
\usepackage[final]{nips_2018}

\usepackage[utf8]{inputenc} 
\usepackage[T1]{fontenc}    
\usepackage{hyperref}       
\usepackage{url}            
\usepackage{booktabs}       
\usepackage{amsfonts}       
\usepackage{nicefrac}       
\usepackage{microtype}      

\usepackage[utf8]{inputenc} 
\usepackage[T1]{fontenc}    
\usepackage{hyperref}       
\usepackage{url}            
\usepackage{booktabs}       
\usepackage{amsfonts}       
\usepackage{nicefrac}       
\usepackage{microtype}      

\usepackage{caption}
\usepackage{subcaption}
\usepackage{times,amssymb,amsmath,amsfonts,openbib,multirow}
\usepackage{graphicx}
\usepackage{xspace}
\usepackage{delarray}
\usepackage{array}
\usepackage{amssymb}
\usepackage{setspace}
\usepackage{comment}
\usepackage{float}
\usepackage{listings,color,tabularx}
\usepackage{afterpage}
\usepackage{wrapfig}
\usepackage{hyperref} 
\usepackage{algorithm}
\usepackage{algorithmic}
\usepackage[skins]{tcolorbox}

\usepackage[
	n,
	operators,
	advantage,
	sets,
	adversary,
	landau,
	probability,
	notions,	
	logic,
	ff,
	mm,
	primitives,
	events,
	complexity,
	asymptotics,
	keys]{cryptocode}
	
\usepackage{csquotes}
\usepackage{fullpage}
\usepackage{dashbox}
\usepackage{todonotes}

\newcommand{\zkp}{\texttt{ZKPoK}}

\DeclareMathOperator{\E}{E}
\DeclareMathOperator{\entropynew}{entropy}
\DeclareMathOperator{\gain}{gain}
\DeclareMathOperator{\tr}{tr}

\setcitestyle{square}

\newcommand{\paragraphe}[1]{\vspace{1ex}\noindent{\em #1} }
\newcommand{\paragraphb}[1]{\vspace{1ex}\noindent{\bf #1} }
\newcommand{\paragraphbe}[1]{\vspace{1ex}\noindent{\bf \em #1} }

\title{Distributed and Secure ML with Self-tallying Multi-party Aggregation}

\author{%
    Yunhui Long \thanks{These two authors contributed equally to the work.} \\
    UIUC \\
    Urbana, Illinois \\
    \texttt{ylong4@illinois.edu} \\
    \And Tanmay Gangwani \footnotemark[1] \\
    UIUC \\
    Urbana, Illinois \\
    \texttt{gangwan2@illinois.edu}
    \And Muhammad Haris Mughees \\
    UIUC \\
    Urbana, Illinois \\
    \texttt{mughees2@illinois.edu} \\
    \And Carl A. Gunter \\
    UIUC \\
    Urbana, Illinois \\
    \texttt{cgunter@illinois.edu} \\
}

\begin{document}
\maketitle

\begin{abstract}
Privacy preserving multi-party computation has many applications in areas such as medicine and online advertisements. In this work, we propose a framework for distributed, secure machine learning among untrusted individuals. The framework consists of two parts: a two-step training protocol based on homomorphic addition and a zero knowledge proof for data validity. By combining these two techniques, our framework provides privacy of per-user data, prevents against a malicious user contributing corrupted data to the shared pool, enables each user to self-compute the results of the algorithm without relying on external trusted third parties, and requires no private channels between groups of users. We show how different ML algorithms such as Latent Dirichlet Allocation, Na\"ive Bayes, Decision Trees etc. fit our framework for distributed, secure computing. 
\end{abstract}
\section{Introduction}
Machine learning models are being increasingly deployed to harness useful information from raw data. Availability of large amounts of training data prevents over-fitting in the models and improves its generalization. However, there is an important tension between the need for large training datasets and the privacy concerns of owners of those datasets. This is best exemplified when considering ML for health and medicine. For instance, assume multiple hospitals, each with access to high-quality (albeit limited in quantity) data about patient medical records. Jointly training a Latent Dirichlet Allocation (LDA) topic model on the union of data would provide insightful information for all the hospitals~\cite{paul2013affects}. However, there is a huge privacy concern for sharing this data as it may contain sensitive information. In this work, we propose a framework for distributed training of ML algorithms among untrusted parties. The framework is secure since the parties can collaboratively train models without revealing their data. Checks for data validity provide robustness against a malicious party contributing illegal data. Furthermore, model aggregation is performed without relying on any external trusted agents. 

\paragraphbe{Related Work.} The problem of distributed and secure machine learning falls under the broad regime of secure multi-party computation (SMPC)~\cite{du2001secure}. Gentry proposed Fully Homomorphic Encryption (FHE)~\cite{gentry2009fully} as a means to achieve SMPC. Current FHE schemes are inefficient and only work with small circuits~\cite{damgaard2012multiparty}. Homomorphism under addition, however, has been extensively studied, and many robust implementations exist~\cite{canny2002collaborative,hao2010anonymous}. 
Hao et al.~\cite{hao2010anonymous} apply additive homomorphism to create an anonymous voting application. Their construction enables self-tallying of votes, precluding the need for trusted third parties for counting. Corrigan et al.~\cite{corrigan2017prio} propose a more scalable secure aggregation protocol and apply it to linear regression. 

\paragraphbe{Contributions.} Our protocol broadens the scope of the ideas presented in~\cite{corrigan2017prio, hao2010anonymous}. Our contribution is three-fold. First, we examine various ML algorithms under the lens of SMPC through homomorphic addition; second, we incorporate input validity checks to dissuade users with malicious data; and third, we propose efficient constructions using basic cryptographic tools like zero-knowledge proofs and ElGamal encryption. We also implement the protocol and present some empirical analysis. 
\section{Distributed and Secure ML}

In the following subsections, we first outline our protocol for secure aggregation of arbitrary integer data vectors from different users. Following that, we detail the reduction of various ML algorithms to generalized vector addition, thereby making them compatible with our framework and enabling secure, distributed training on aggregated data.

\subsection{Threat Model and Notations}
Suppose there are $n$ users. Each user $U_i$ owns an integer data vector $T_i$ of size $m$. We then desire the output of the vector addition $T = \sum_{i=1}^n T_i$ with the following properties:
\begin{itemize}
    \item {\em Privacy:} The contents of $T_i$ should be kept a secret from users other than $i$. In our protocol, this secrecy is maintained unless all of the other users have been compromised.
    \item {\em Input validity:} A malicious user should not be able to corrupt $T$ by providing {\em unexpected} values. Depending on the ML algorithm, this could mean preventing a large integer input which can disturb cumulative statistics, or a negative input for an always-positive variable.
    \item {\em Self-tallying:} Any user should be able to compute $T$ without relying on external talliers.
    \item {\em No private channels:} We assume only the availability of a publicly verifiable ledger (e.g. blockchain) and no user-to-user private channels. This offers dispute-freeness. 
\end{itemize}

The zero-knowledge proof-of-knowledge (\zkp) used in our protocol are expressed in Camenisch-Stadler notation~\citep{camenisch1997efficient}: $\zkp_x{(w) : L(w, x)}$. Here, $x$ is the public statement, $w$ is the secret witness and $L$ represents the conditions that the statement and witness must satisfy. 

\subsection{Two-round Protocol for Homomorphic Vector Addition}
\label{subsec:two-round-pro}
Let $C$ be a finite cyclic group of prime order $q$ in which the discrete log problem is hard, and $g$ be a generator in $C$. There are $n$ users, each with a secret key $\sk_i$, and they agree on $(C, g)$. User $U_i$'s contribution to the aggregate ($T$) is a $m$-dimensional vector ($T_i$).

\noindent
{\textbf{First Round: }}Each user $U_i$ selects $m$ random values ($x_{i1}, x_{i2}, \dots, x_{im}) \in_R \mathbb{Z}_q$, publishes to the public ledger the values $(g^{x_{i1}}, g^{x_{i2}}, \dots, g^{x_{im}})$ and a \zkp\ of discrete log ($\zkp_A(a) : g^a = A$) for each $x_{ij}$ $(1 \le j \le m)$. At the end of this round, each $U_i$ checks the validity of 
the \zkp s on the ledger, and computes:
\begin{equation*}
\label{eq:h_ij}
    h_{ij} = g^{y_{ij}} = \frac{\prod_{k=1}^{i-1} g^{x_{kj}}}{\prod_{k=i+1}^{n} g^{x_{kj}}}, \quad \forall  1 \le j \le m.
\end{equation*}

\noindent
{\textbf{Second Round: }} Each user $U_i$ computes the ElGamal encryption of $T_{ij}$ for $1 \le j \le m$ as 
\begin{equation*}
\label{eq:enc_G}
\E[T_{ij}] = (g^{x_{ij}}, g^{T_{ij}}h_{ij}^{x_{ij}}).
\end{equation*}
$U_i$ then publishes the encrypted vector $(\E[T_{i1}], \E[T_{i2}], \dots, \E[T_{im}])$. Our construction of the public keys ($h_{ij}$) is similar to that in the first round of anonymous voting in~\cite{hao2010anonymous}. Hence, it follows that by multiplying the correct ciphertext values, any user can compute $g^{\sum_i T_{ij}}$ for $1 \le j \le m$. Although computing $\sum_i T_{ij}$ requires taking a discrete log, the range of $\sum_i T_{ij}$ is generally not large, and a baby-step/giant-step approach~\cite{lenstraalgorithms} is practical. At the end of the second round, each user can produce the vector summation $T$ by self-tallying the values for each index ($j$) of the vector.

To discourage malicious users from submitting encryptions of 
corrupted (or disallowed) $T_i$ in the second round, we augment the protocol with input validity checks. Specifically, along with the encrypted $T_i$, each user is required to submit another proof to the ledger which can be validated by others for compliance of the input data. We consider two such compliance conditions - $L^2$-norm and $L^1$-norm of $T_i$. In many algorithms, such as collaborative filtering (Appendix~\ref{apps:cf}), imposing a bound on $L^2$-norm, i.e. $\norm{T_i}_2$, serves as a reasonable precondition. In notation, we want the \zkp:

\begin{equation}\label{eq:range_l2}
    \zkp_{(\vec{x},\vec{y}, B)}(\vec{a}, \vec{r}) : (x_i,y_i) = (g^{r_i}, h^{r_i} \cdot g^{a_i}) \land \norm{\vec{a}}_2 \le B
\end{equation}

Bounding the $L^2$-norm does not guarantee that all (or any) of the entries in the vector $T_i$ are non-negative. Non-negative inputs are required in some algorithms like LDA and decision trees (Appendix~\ref{apps:dt}). Moreover, it is more useful to bound the $L^1$-norm, i.e. $\norm{T_i}_1$, than the $L^2$-norm:

\begin{equation}\label{eq:range_l1}
    \zkp_{(\vec{x},\vec{y}, B)} (\vec{a}, \vec{r}) : (x_i,y_i) = (g^{r_i}, h^{r_i} \cdot g^{a_i}) \land \norm{\vec{a}}_1 \le B \land a_i \ge 0
\end{equation}

The \zkp s in Equations (\ref{eq:range_l2}) and (\ref{eq:range_l1}) are constructed from other simpler \zkp s mentioned in Appendix~\ref{appx:zkp}. We deem this construction to be an important contribution of this work. It is detailed in Appendix~\ref{appn:range_proofs}, along with the complete steps run by the prover and the verifier to generate and validate the proofs. We also mention future work on optimizing these proofs.

\subsection{Reduction of Algorithms to Vector Addition}
\label{subsec:apps}

We now discuss several algorithms which fit into our framework for distributed and secure computation. In each case, it can be shown that the algorithm decomposes into a simple addition of integer vectors (or matrices) created from disjoint data pieces. This enables the various untrusting parties to safely engage in joint training of ML models using the protocol from previous subsection. Table~\ref{table:appssummary} summarizes the algorithms, along with a validity check ($L^1$, $L^2$-norm) for it, and the significance of the check. Note that the $L^1$-norm bound check (Eq.~\ref{eq:range_l1}) also includes the non-negativity constraint. We explain one algorithm (LDA) in detail here; reduction of other algorithms is in Appendix~\ref{subsec:ml}. 

\begin{table}[h]
\centering
\begin{tabular}{lll}
\hline \hline
\textbf{Application} & \textbf{Validity} & \begin{tabular}[c]{@{}l@{}} \textbf{Significance of Check}\end{tabular} \\ \hline \hline
\begin{tabular}[c]{@{}l@{}}Latent Dirichlet Allocation \end{tabular} & $L^1$-norm & \begin{tabular}[c]{@{}l@{}}Limit number of times a word is assigned to \\ a topic by each user; disallow negative values\end{tabular} \\ \hline
Decision Trees & $L^1$-norm &
\begin{tabular}[c]{@{}l@{}} Limit number of training samples per user; \\ disallow negative values \end{tabular} \\ \hline
Na\"ive Bayes & $L^1$-norm & 
\begin{tabular}[c]{@{}l@{}} Limit number of training samples per user; \\ disallow negative values \end{tabular} \\ \hline
Cumulative Voting & $L^1$-norm & 
\begin{tabular}[c]{@{}l@{}} Limit total number of votes by each voter; \\ disallow negative values \end{tabular} \\ \hline
Linear Regression & $L^2$-norm & Limit contribution to $\beta$, prevent over-fitting \\ \hline
\begin{tabular}[c]{@{}l@{}}Collaborative Filtering\end{tabular} & $L^2$-norm & \begin{tabular}[c]{@{}l@{}}Limit contribution to the preference matrix\end{tabular} \\ \hline
\end{tabular}
\vspace{5mm}
\caption{Summary of Algorithms}
\vspace{-5mm}
\label{table:appssummary}
\end{table}

\paragraphb{Latent Dirichlet Allocation.}
%
LDA is a generative probabilistic modelling technique for collections of discrete data such a text documents~\cite{blei2003latent}. For each document $j$, there is a multinomial distribution  $\theta_j$ over K hidden topics. Also, the $k^{th}$ topic is represented by a multinomial distribution $\phi_k$ over the word vocabulary. $x_{ij}$, which is the $i^{th}$ word in document $j$, is associated with a latent topic assignment $z_{ij}$. Given all words in all documents $\bold{x} = \{x_{ij}\}$, the inference task in LDA is to compute the posterior over $\bold{z} = \{z_{ij}\}$, $\theta_j$ and $\phi_k$. 

We summarize the approximate distribued LDA algorithm proposed by Newman et al.~\cite{newman2009distributed} which uses collapsed Gibbs sampling to sample the posterior $z_{ij}$ at each state of the Markov chain. The algorithm initially divides the document corpus among different processors. We consider different processors as different users. Each user does local Gibbs sampling for a few iterations before synchronizing with other users. We encourage interested readers to look at Algorithm 1. in~\cite{newman2009distributed}. The synchronization involves a matrix reduction operation and is the only medium through which the privacy of a user's data could be violated:
$N_{wk} \leftarrow \sum_{u \in \textrm{users}} N_{wk}^{(u)}$.

\paragraphe{Computing $\mathbf{N_{wk}}$:} After local Gibbs sampling for few iterations, each user computes $N_{wk}^{(u)}$, which is a matrix containing counts of the number of times a particular word is assigned to a particular topic. The encrypted matrix from each user can be homomorphically added and the result $N_{wk}$ can be obtained by each user independently by self-tallying. To prevent a malicious user from including large or negative values in $N_{wk}^{(u)}$, the parties can decide on a bound for the $L^1$-norm of the input, and require that each user provide the corresponding range proofs.
\section{Implementation}

In this section, we evaluate the homomorphic vector addition protocol through the application of cumulative voting, and summarize some observations. In cumulative voting, each voter is given $B-1$ number of votes, and can arbitrarily distribute these votes among the candidates. A voter's input is considered legal as long as the total number of votes given by her is less than $B$. The voters are allowed to vote for more than one candidate and to put more than one vote on preferred candidates. Suppose there are $n$ voters and $m$ candidates. Let the vector $T_i = (T_{i1}, T_{i2}, \dots, T_{im}) \quad (1 \le i \le n)$ be the votes of voter $i$, where $T_{ij}$ is the number of votes given by voter $i$ to candidate $j$. The result of cumulative voting can be tallied by adding the vote vectors from all voters:
\begin{equation*}
    T_{\textrm{result}} = (\sum_{i=1}^{n}T_{i1}, \sum_{i=1}^{n}T_{i2}, \dots, \sum_{i=1}^{n}T_{im}). 
\end{equation*}
To guarantee the fairness of cumulatve voting, it is necessary for each user to provide a \zkp ~for $L^1$-norm bound on each voting vector $T_i$. This limits the total number of votes by each voter and disallows negative votes (Table~\ref{table:appssummary}).

\begin{wrapfigure}[17]{r}{0.6\textwidth}
\vspace{-5mm}
\centering
  \includegraphics[width=0.6\textwidth]{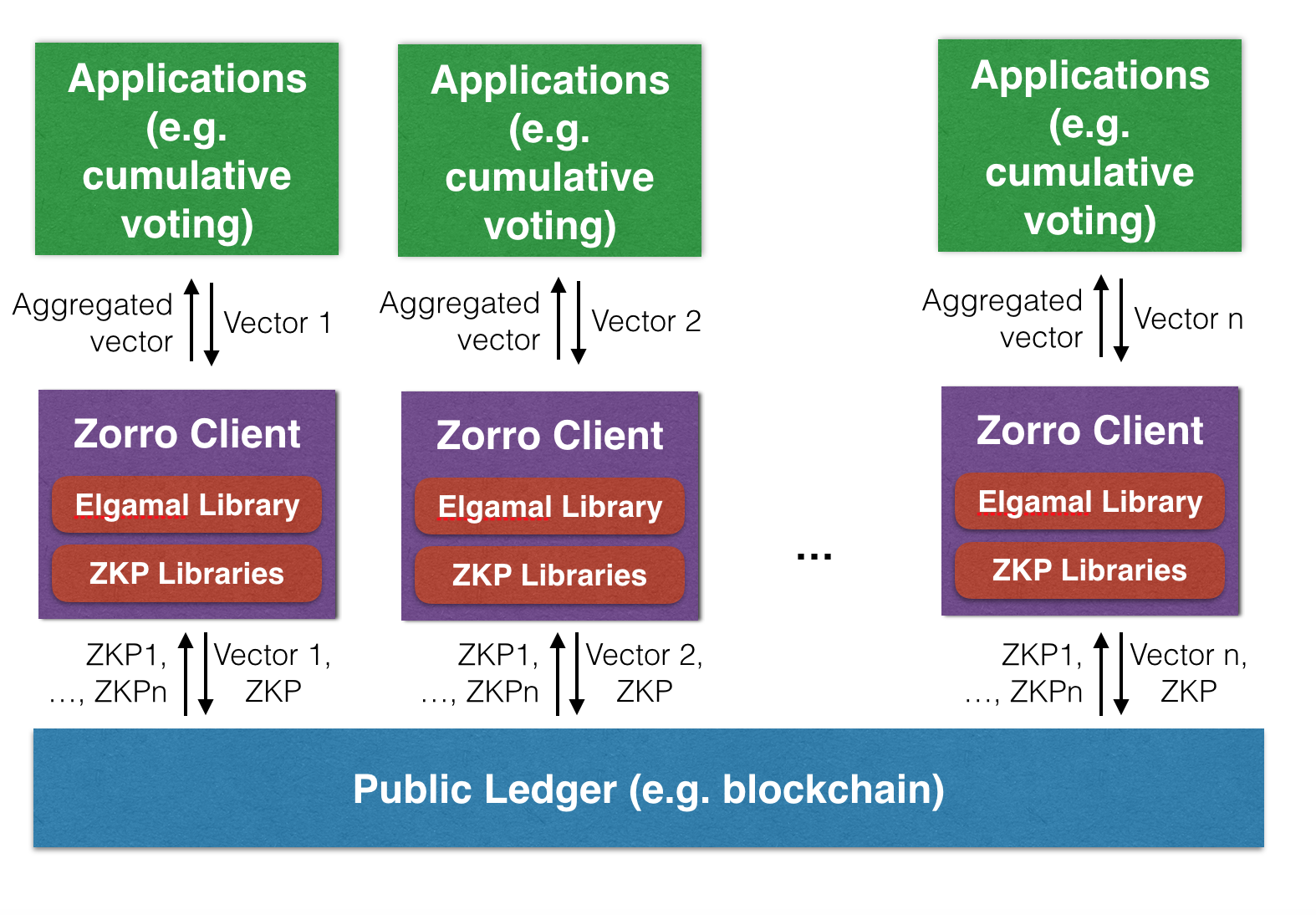}
  \vspace{-5mm}
  \caption{Structure of Implementation}
  \label{fig:struct}
\vspace{-5mm}
\end{wrapfigure}

Our implementation\footnote{https://github.com/tgangwani/Zorro\_SMPC} consists of the following layers: an ElGamal Encryption library implemented over elliptic curves, \zkp ~libraries, an interfacing client, which we call \textbf{Zorro client}, and a cumulative voting application (Figure~\ref{fig:struct}). To simulate the environment of the blockchain, we implement a
  public ledger class that stores the encrypted data and \zkp s. In practice, the public ledger can be replaced by a smart contract and deployed on the Ethereum block chain. Further details on the components of the implementation are in Appendix~\ref{appx:eval}. Therein, we also include an analysis on the machine time taken to generate and verify the \zkp s. The computational cost for \zkp s depends on the vector length (total candidates) and the bound (maximum votes allowed per voter), with the former being the more dominant factor. We provide some discussion on the
time-complexity of baby-step/giant-approach~\cite{lenstraalgorithms}, showing that it speeds up the discrete-log step. We also measure the effects of using integer precision rather than floating point precision for a simple linear regression problem, concluding that the accuracy-loss can be controlled. A more extensive study is interesting future work.
\vspace{-2mm}
\section{Conclusion}
\vspace{-2mm}
In this paper, we outline a protocol for secure, distributed computing with multiple mutually distrusting parties. It includes input validity checks (bound on $L^1$-norm and $L^2$-norm) to guard against malicious users. It uses efficient constructions to prove information in zero-knowledge, uses a public-ledger to offer dispute-freeness, and is self-tallying, thus obviating presence of trusted third parties. We show how popular ML algorithms such as LDA, Na\"ive Bayes, Decision Trees etc. can be used with our framework. Furthermore, we implement our protocol on top of cryptographic constructs and open-source our Zorro client for multi-party cumulative voting.

\small{
\bibliographystyle{abbrv}
\bibliography{reference.bib}}

\newpage

\section{Appendix}


\subsection{Zero Knowledge Proof-of-knowledge}
\label{appx:zkp}

We express the various zero-knowledge proof-of-knowledge (\zkp) used in our protocol in Camenisch-Stadler notation: 
\begin{center}
   $\zkp_x{(w) : L(w, x)}$,
\end{center}
where $x$ is the public statement, $w$ is the secret witness and $L$ represents the conditions that the statement and witness must satisfy. We use the following \zkp s:
\begin{gather}
\zkp_A(a) : g^a = A \label{zkpeq1} \\
\zkp_{(g,h,u,v)}(w) : g^w = u \land h^w = v \label{zkpeq2}\\
\zkp_{(x,y)}(r) : (x,y) = (g^r,h^r)  \lor (x,y) = (g^r, h^r \cdot g) \label{zkpeq3} \\ 
\zkp_{(x_a,y_a,x_b,y_b)}(a,b,r_a,r_b) : (x_a,y_a) = (g^{r_a},h^{r_a} \cdot g^a) \land (x_b,y_b) = (g^{r_b},h^{r_b} \cdot g^b) \land b = a^2 \label{zkpeq4} 
\end{gather}
In words, (1) is the \zkp\ of discrete log; (2) proves that $(g,h,u,v)$ forms a Diffie-Hellman 4-tuple~\cite{hazay2010efficient}; (3) is \zkp\ for ElGamal encryption of $m \in \{0,1\}$; (4) proves the square relationship between pre-images of two ElGamal encryptions. In our implementation, we make them non-interactive by using Fiat-Shamir's heuristics~\cite{fiat1986prove}.

\paragraph{Proof of Discrete Log, Eq.~\ref{zkpeq1}}
\begin{pcvstack}[center]
\fbox{
\pseudocode{%
\textbf{ Prover}(a,A = g^a) \< \< \textbf{ Verifier}(A)  \\[0.1\baselineskip][\hline]
 \<\< \\[-0.5\baselineskip]
 \centering
k \sample \mathbb{Z}_q  \< \< \\
K:= g^k  \< \sendmessageright*{K} \< c\sample \mathbb{Z}_q  \\
s:= k +ca  \< \sendmessageleft*{c} \< \\
  \< \sendmessageright*{s} \< g^s \stackrel{?}{=} KA^c\\
 }
 }
\end{pcvstack}


\paragraph{Proof of Diffie-Hellman Tuple, Eq.~\ref{zkpeq2}}

\begin{pcvstack}[center]
\fbox{
\pseudocode{%
\textbf{Prover}(G,q,g,h,u,v)\<\< \textbf{Verifier}(G,q,g,h,u,v)
 \\[0.1\baselineskip][\hline]
 \<\< \\[-0.5\baselineskip]
 \centering
 w \mid u=g^w,v=h^w\<\<\\
 r\sample Z_q \mid a=g^r, b=h^r \<\sendmessageright*{a,b}\<e \sample \{0,1\}^t\mid 2^t<q\\ 
z=r+ew(\mod q)\<\sendmessageleft*{e}\< \\
\<\sendmessageright*{e}\< g^z \stackrel{?}{=} au^e,h^z \stackrel{?}{=} bv^e \\}
}
\end{pcvstack}


\paragraph{Proof of encryption of $\mathbf {x_j \in \{0,1\}}$, Eq.~\ref{zkpeq3}}

\begin{pcvstack}[center]
\fbox{
\pseudocode{\textbf{ Prover} \< \< \textbf{ Verifier}  \\[0.1\baselineskip][\hline]
 \<\< \\[-0.5\baselineskip]
 \centering
w,r_1,d_1\in_R Z_q |& w,r_2,d_2 \in_R Z_q \< \< \\
 x \gets g^{x_j}&x \gets g^{x_j} \<\< \\
 a_1 \gets g^{r_1}x^{d_1}& a_1 \gets g^{w} \<\< \\
 b_1 \gets h^{r_1}y^{d_1}& b_1 \gets h^{w} \<\< \\
a_2 \gets g^{w}& a_2 \gets g^{r_2}x^{d_2} \<\< \\
b_2 \gets h^{w}& b_2 \gets h^{r_2}(y/g)^{d_2}\< \sendmessageright*{x,y,a_1,b_1,a_2,b_2}\<  c \sample \mathbb{Z}_q \\ 
d_2\gets c- d_1 & d_1 \gets c- d_2\< \sendmessageleft*{c} \< c\stackrel{?}{=} d_1+d_2 \\
r_2 \gets w -x_j d_2 & r_1 \gets w- x_jd_1 \<  \sendmessageright*{d_1,d_2,r_1,r_2} \< a_1\stackrel{?}{=} g^{r_1}x^{d_1}\\
\hspace{1em} & \< \< b_1\stackrel{?}{=} h^{r_1}y^{d_1}\\
\hspace{1em} & \< \< a_2\stackrel{?}{=} g^{r_2}x^{d_2}\\
\hspace{1em} & \< \< b_1\stackrel{?}{=} h^{r_1}y^{d_1}\\}
}
\end{pcvstack}


\paragraph{Proof of square relation, Eq.~\ref{zkpeq4}}

\begin{pcvstack}[center]
\fbox{
\pseudocode{%
\textbf{ Prover}(s_a,s_b \in_R \mathbb{Z}_q ) \< \< \textbf{ Verifier}  \\[0.1\baselineskip][\hline]
 \<\< \\[-0.5\baselineskip]
 \centering
 A \gets (g^{s_a},\gamma^a h^{s_a})\mod p \<\< \\ 
 B \gets (g^{s_b},\gamma^b h^{s_b})\mod p \<\< \\
 x,r_a,r_b \sample \mathbb{Z}_q  \<\< \\ 
 C_a \gets (g^{r_a},\gamma^x h^{r_a}) C_b \gets A^x(g^{r_b},h^{r_b}) \<\sendmessageright*{C_a,C_b}\< c \sample \mathbb{Z}_q \\
v \gets ca+x(\mod q) \<\sendmessageleft*{c}\< \\
z_a \gets cs_a + r_a(\mod q)\<\< \\
z_b = c(s_b   as_a) + r_b(\mod q)\<\sendmessageright*{v,z_a,z_b}\< (g^{z_a}, \gamma^vh^{z_a}) \stackrel{?}{=} A^cC_a \\
\<\< A^v(g,h)^{z_b} \stackrel{?}{=} B^cC_b \\
 }
 }
\end{pcvstack}

\clearpage

\subsection{Range Proofs}
\label{appn:range_proofs}
\subsubsection{Range-proof for $L^2$-norm}

\label{sub:zkp-l2}
\renewcommand{\norm}[1]{\left\lVert#1\right\rVert}
%
In notation, we want the \zkp:
\begin{equation*}
    \zkp_{(\vec{x},\vec{y}, B)}(\vec{a}, \vec{r}) : (x_i,y_i) = (g^{r_i}, h^{r_i} \cdot g^{a_i}) \land \norm{\vec{a}}_2 \le B
\end{equation*}

\noindent
We now show how we compose this \zkp ~using the basic \zkp s (Eq.~\ref{zkpeq1}-~\ref{zkpeq4}).

\noindent
{\textbf{Step 1: }}
Each user $U_i$ generates an ElGamal public key ($h_i$) from its private key $sk_i$, and encrypt each $T_{ij}$ as: 
\begin{equation}
\label{eq:enc_G*}
\E^*[T_{ij}] = (g^{x_{ij}}, g^{T_{ij}}h_i^{x_{ij}}).
\end{equation}
Then, $U_i$ proves that $\E[T_{ij}]$ and $\E^*[T_{ij}]$ encrypt the same plaintext. For this, it's sufficient to prove that $(g, \frac{h_{ij}}{h_i}, g^{x_{ij}}, \frac{\E[T_{ij}]}{\E^*[T_{ij}]})$ is a Diffie-Hellman 4-tuple using \zkp\ (Eq.~\ref{zkpeq2}). This is because of the following equation:
\begin{equation*}
\frac{\E[T_{ij}]}{\E^*[T_{ij}]} = (1, (\frac{h_{ij}}{h_i})^{x_{ij}}).
\end{equation*}

\noindent
{\textbf{Step 2: }}
Each user $U_i$ calculates the square vector ($\mathbf{w}_i$), encrypts it using the ephemeral key detailed below in Eq.~\ref{eq:enc_w}, and publishes the encryption on the public ledger. It also provides a proof of the square relation (\zkp\ (Eq.~\ref{zkpeq4}))
\begin{equation*}
\mathbf{w}_i = (w_{i1}, w_{i_2}, \dots, w_{im}) = (T_{i1}^2, T_{i2}^2, \dots, T_{im}^2).
\end{equation*}

\noindent
Let $B$ be the bound on $\norm{T_i}_2$. $U_i$ needs to prove the following: 
\begin{equation*}
s = \sum_{j=1}^{m} w_{ij} < B^2.
\end{equation*}

\noindent
We provide a range-proof for $s$ by decomposing $s$ into binary representations~\cite{canny2002collaborative}. Let $L = 2\log_2{B}$. Then, $s$ can be represented by an $L$-digit binary value, and expressed as a weighted sum of each digit:
\begin{equation*}
s = \sum_{l=0}^{L-1} 2^l s_{il}.
\end{equation*}

\noindent
To prove that ${\norm{T_i}_2^2}<B^2$, we need two sub-proofs. Firstly, we need to show that $s_{il} \in \{0,1\}$ for all $0 \le l \le L-1$. This can be easily done by \zkp\ (Eq.~\ref{zkpeq3}). The second challenge is to prove that each $s_{il}$ is indeed a digit in the binary representation of $\sum_{j=1}^{m} w_{ij}$. That is, the user should show the following, without revealing the values of $w_{ij}$ and $s_{ij}$:
\begin{equation}
\label{eq:binary_sum}
\sum_{j=1}^{m} w_{ij} = \sum_{l=0}^{L-1} 2^l s_{il}.
\end{equation}

\noindent
The protocol to validate Eq.~\ref{eq:binary_sum} is as follows:\\
First, each user $U_i$ selects $L$ random values $x'_{i1}, x'_{i2}, \dots, x'_{iL} \in_R \mathbb{Z}_q$, and encrypts each $s_{il}$ $0 \le l \le L-1$ as:
\begin{equation}
\label{eq:enc_s}
    \E[s_{il}] = (g^{x'_{i(l+1)}}, g^{s_{il}}h_{i}^{x'_{i(l+1)}}).
\end{equation}

\noindent
Then, each user $U_i$ selects $m$ random values $x^*_{i1}, x^*_{i2}, \dots, x^*_{im} \in_R \mathbb{Z}_q$. For all $1 \le j \le m$, $U_i$ calculates:
\begin{equation}
\label{eq:r}
    r_{ij} = (\sum_{k=1}^{j-1} x^*_{ik} - \sum_{k=j+1}^{m} x^*_{ik})x^*_{ij}.
\end{equation}

\noindent
Assuming $m > L$, $U_i$ encrypts each $w_{ij}$ $(1 \le j \le m)$ as:
\begin{equation}
\label{eq:enc_w}
    \E[w_{ij}] = 
\begin{cases}
\begin{aligned}
    &(g^{r_{ij}+x'_{ij}2^{(j-1)}}, g^{w_{ij}}h_{i}^{r_{ij}+x'_{ij}2^{(j-1)}}) \qquad & \textrm{if} \quad j \le L \\
    &(g^{r_{ij}}, g^{w_{ij}}h_{i}^{r_{ij}}) \qquad & \textrm{if} \quad j > L
\end{aligned}
\end{cases}.
\end{equation}

\noindent
To verify Eq.~\ref{eq:binary_sum}, a verifier needs to check that: 
\begin{equation*}
     \prod_{j=1}^{m}\E[w_{ij}] = \prod_{l=0}^{L-1}\E[s_{il}]^{2^l}.
\end{equation*}
Or equivalently,  
\begin{equation}
\label{eq:check}
\begin{cases}
\begin{aligned}
    &\prod_{j=1}^{L}g^{r_{ij}+x'_{ij}2^{(j-1)}} \prod_{j=L+1}^{m} g^{r_{ij}} = \prod_{l=0}^{L-1} g^{x'_{i(l+1)}2^{l}} \\
    &\prod_{j=1}^{L} g^{w_{ij}}h_{i}^{r_{ij}+x'_{ij}2^{(j-1)}} \prod_{j=L+1}^{m} g^{w_{ij}}h_{i}^{r_{ij}} = \prod_{l=0}^{L-1} g^{s_{il}2^{l}}h_{i}^{x'_{i(l+1)}2^{l}}
\end{aligned}
\end{cases}.
\end{equation}
Since $\sum_{j=1}^{m}r_{ij} = 0$, the noise terms $r_{ij}$ cancels out. Eq.~\ref{eq:check} should hold if and only if $\sum_{j=1}^{m} w_{ij} = \sum_{l=0}^{L-1} 2^l s_{il}$, thereby completing the proof for Eq.~\ref{eq:binary_sum}. An alternative to using encryptions where the noise terms $r_{ij}$ nullify each other is to use a Diffie-Hellman proof (\zkp\ (Eq.~\ref{zkpeq2})) for Eq.~\ref{eq:binary_sum}. It achieves the same goal, albeit at the cost of an extra \zkp. Below we summarize the complete steps run by the prover and the verifier to generate and validate the range proof for $L^2$-norm, respectively. \\

\noindent
As mentioned previously, $U_i$ provides a proof that $w_{ij} = T_{ij}^2$, for all $1 \le j \le m$ (\zkp\ (Eq.~\ref{zkpeq4})). We use the construction by Canny~\cite{canny2002collaborative} for this \zkp. Canny's proof requires that $w_{ij}$ and $T_{ij}$ to be encrypted under exponential ElGamal encryption with the {\em same} public key. Therefore, in the proof, we use $\E[w_{ij}]$ and $\E^*[T_{ij}]$, which are both encrypted under the same key $h_i$.

\let\AND\undefined
\let\OR\undefined
\let\NOT\undefined

\begin{algorithm}[H]
\caption{Proof generation by user $i$}
\label{algo:ZKP-prover}
\begin{algorithmic}[1]
   	\REQUIRE $(T_{i1}, T_{i2}, \dots, T_{im})$, $(\E[T_{i1}], \E[T_{i2}], \dots , \E[T_{im}])$, \\
   	$(h_{i1}, h_{i2}, \dots, h_{im})$, $(x_{i1}, x_{i2}, \dots, x_{im})$, \\
   	ElGamal parameters $(g,h_i)$, $L = 2\log_2{B}$
   	\STATE Encrypt each $T_{ij}$ as $\E^*[T_{ij}]$ (Eq.~\ref{eq:enc_G*})
   	\STATE For each $T_{ij}$, generate proof for Diffie-Hellman 4-tuple $(g, \frac{h_{ij}}{h_i}, g^{x_{ij}}, \frac{\E[T_{ij}]}{\E^*[T_{ij}]})$ 
	\STATE Calculate $w_{ij} = T_{ij}^2$, and $s = \sum_{j=1}^{m}w_{ij}$
	\STATE Calculate $s_{i0}, s_{i1}, \dots, s_{i(L-1)}$ such that $s = \sum_{l=0}^{L-1}s_{il}2^{l}$   	
   	\STATE Generate $L$ random values $x'_{i1}, x'_{i2}, \dots, x'_{iL} \in_R \mathbb{Z}_q$
   	\STATE Encrypt each $s_{il}$ as $\E[s_{il}]$ (Eq.~\ref{eq:enc_s})
   	\STATE Generate proof for $s_{il} \in \{0,1\}$, for each $s_{il}$
   	\STATE Generate $m$ random values $x^*_{i1}, x^*_{i2}, \dots, x^*_{im} \in_R \mathbb{Z}_q$
   	\STATE Calculate $r_{ij}$ (Eq.~\ref{eq:r}) and encrypt each $w_{ij}$ as $\E[w_{ij}]$ (Eq.~\ref{eq:enc_w})
    \STATE Generate proof for $(w_{ij}=T_{ij}^2)$, for each $w_{ij}$
    \STATE Send the following messages to the verifier: \\
    $(\E[T_{i1}], \E[T_{i2}], \dots , \E[T_{im}])$, $(\E^*[T_{i1}], \E^*[T_{i2}], \dots , \E^*[T_{im}])$, \\
    $(h_{i1}, h_{i2}, \dots, h_{im})$, $h_i$, \zkp$(g, \frac{h_{ij}}{h_i}, g^{x_{ij}}, \frac{\E[T_{ij}]}{\E^*[T_{ij}]})$ for each $T_{ij}$, \\
    $(\E[s_{i0}], \E[s_{i1}], \dots, \E[s_{i(L-1)}])$, $(\E[w_{i1}], \E[w_{i2}], \dots, \E[w_{im}])$,\\
    \zkp$(s_{il} \in \{0,1\})$ for each $s_{il}$, \zkp$(w_{ij}=T_{ij}^2)$ for each $w_{ij}$
\end{algorithmic}
\end{algorithm}

\begin{algorithm}[H]
\caption{Proof verification}
\label{algo:ZKP-verifier}
\begin{algorithmic}[1]
\REQUIRE Messages received from user $i$ in Algorithm~\ref{algo:ZKP-prover}
\STATE Verify \zkp$(g, \frac{h_{ij}}{h_i}, g^{x_{ij}}, \frac{\E[T_{ij}]}{\E^*[T_{ij}]})$ for each $T_{ij}$
\STATE Verify $\prod_{j=1}^{m}\E[w_{ij}] = \prod_{l=0}^{L-1}\E[s_{il}]^{2^l}$
\STATE Verify \zkp$(s_{il} \in \{0,1\})$ for each $s_{il}$
\STATE Verify \zkp$(w_{ij}=T_{ij}^2)$ for each $w_{ij}$
\end{algorithmic}
\end{algorithm}
\subsubsection{Range-proof for $L^1$-norm}
\label{sub:zkp-l1}
%
In notation, we want the \zkp:
\begin{equation*}
    \zkp_{(\vec{x},\vec{y}, B)} (\vec{a}, \vec{r}) : (x_i,y_i) = (g^{r_i}, h^{r_i} \cdot g^{a_i}) \land \norm{\vec{a}}_1 \le B \land a_i \ge 0
\end{equation*}

With slight abuse of terminology, we'll call this proof as range-proof for $L^1$-norm, although it is much stronger and includes the additional proof for non-negativity of values. The proof proceeds in a manner very similar to section~\ref{sub:zkp-l2}, but we now require a range-proof for each element ($T_{ij}$) of the vector $T_i$. Like before, we do this by decomposing $T_{ij}$ into binary representations~\cite{canny2002collaborative}.

\subsubsection{Optimizations}
\label{subsubsec:opt}
The range-proof for $L^1$-norm of a vector requires range-proofs for all the elements of the vector. This leads to large time and space overheads in practice. There are a few approaches in literature which we can use to overcome this. Camenisch et al.~\cite{camenisch2008efficient} use a base $B, (B > 2)$ decomposition of a number $s$ rather than base 2. This reduces the number of ciphertexts sent from the prover to the receiver. The authors use an elegant protocol to prove set membership $s_i \in \phi=\{0, \dotsc, B-1\}$. The basic idea is to have the verifier provide a signature on each element of the set $\phi$. The prover then proves in zero knowledge that it possesses a signature on the committed value $s_i$. The proof is sound because the prover can't fake a signature on a value outside the set $\phi$. The efficiency of the protocol stems from the fact that the same set of signatures from the verifier can be used multiple times to commit to different $s_i$ values.

Peng et al.~\cite{peng2010batch} propose an approach called {\em batched} range proofs to improve computational efficiency. They also use a higher base decomposition and reduce the problem to proof of membership in a set of size $k$. Set membership is proved using a proof of knowledge of 1-out-of-$k$ discrete logarithms. The novelty of their protocol is in batching (or combining) $n$ such instances of 1-out-of-$k$ discrete logarithms proof into one single proof, using generalized Pedersen commitments. This reduces the complexity of the overall protocol.

\clearpage

\subsection{Reduction of ML Algorithms to Vector Addition}
\label{subsec:ml}

\subsubsection{Decision Trees}
\label{apps:dt}
Decision Trees are widely used for non-linear multi-class classification. The ID3 algorithm~\cite{id3} for decision trees forms the tree by a recursive process. In each step of the recursion, a metric known as {\em entropy gain} is calculated for each feature in the feature-vector using the data-set available in the step. The feature with the highest entropy gain is selected as the root of the ensuing sub-tree. The recursion is usually terminated after a short depth to prevent over-fitting, with the leaves of the tree forming the class labels. 

In the equations below, $D$ is the complete dataset and $q_j$ is the fraction of samples with label $j$ in $D$. Let $f$ be any feature which takes values $v \in F$. $D_v$ is the set of samples from $D$ where the feature $f$ has a value $v$.

\begin{gather*}
    \entropynew(D) = - \sum_j q_j \log q_j \\
    \gain(f) = \entropynew(D) - \sum_{v \in F} \frac{|D_v|}{|D|} \entropynew(D_v) 
\end{gather*}

\noindent
\textbf{Computing $\mathbf{entropy(D)}$:} Let $D_i$ be the fraction of the complete dataset in possession of user $i$. If the total number of labels is $k$, each user creates an encrypted vector $(c_1, \dotsc, c_k)$, where $c_j$ is the number of samples of label $j$ in $D_i$. To prevent a malicious user from supplying large values for $c_j$ which can corrupt the model parameters, range proofs for $c_j$ and $L^1$-norm of the vector are required. Each user can then calculate $q_j$, and hence $\entropynew(D)$, by homomorphically adding all the vectors.\\

\noindent
\textbf{Computing $\mathbf{gain(f)}$:} For ease of exposition, assume that $F = \{0,1\}$, and there is only one feature $f$. User $i$ creates two encrypted vectors $(p_1, \dotsc, p_k)$ and $(q_1, \dotsc, q_k)$, where $p_j$ is the number of samples in $D_i$ with $\{f=0, label=j\}$, and $q_j$ is the number of samples in $D_i$ with $\{f=1, label=j\}$. For input validity, a proof for $c_j = p_j + q_j$ is required. As before, using homomorphic addition, each user can compute $\entropynew(D_v)$, and hence $\gain(f)$.

\subsubsection{Na\"ive Bayes}
\newcommand{\argmax}{\arg\!\max}
Na\"ive Bayes classifiers are probabilistic classifiers which utilize the {\em na\"ive} assumption of conditional independence of the features, given the class label. Given a data sample $(x_1, \dotsc, x_n)$, it uses Bayes' theorem to calculate the likelihood that the sample belongs to a particular class label:

\begin{equation*}
    \Pr(y | x_1, \dotsc, x_n) = \frac{\Pr(y)\Pr(x_1, \dotsc, x_n | y)}{\Pr(x_1, \dotsc, x_n)}.
\end{equation*}
     
Using Na\"ive Bayes assumption and simplifying, the classification rules is given by-

\begin{equation*}
    \hat{y} = \argmax_y \Pr(y) \prod_{i=1}^{n} \Pr(x_i|y).  
\end{equation*}
    
The model parameters that are learned from the training data are $\Pr(y)$ and $\Pr(x_i|y)$. 
Although different assumptions can be made on the distribution of the parameters, we estimate them empirically using the counts from the training data: 

\begin{gather*}
    \Pr(y = l) = \frac{|y=l|}{|D|}, \\
    \Pr(x_i = m|y=l) = \frac{|x_i = m, y = l|}{|y=l|}.
\end{gather*}

\noindent
\textbf{Computing $\mathbf{Pr(y=l)}$:} Identical to the computation of $q_j$ in ID3. Each user contributes a vector $(c_1, \dotsc, c_k)$, along with range proofs. \\

\noindent
\textbf{Computing $\mathbf{Pr(x_i = m|y=l)}$:} Identical to the computation of $\entropynew(D_v)$ in ID3. Each user creates as many vectors as the number of possible values for $x_i$, along with a proof that the vectors sum to $(c_1, \dotsc, c_k)$.

\subsubsection{Linear Regression}
\label{apps:lr}
Given data samples of the form $(\vec{x}, y)$, linear regression models $y$, which is referred to as the dependent variable, as a linear combination of $\vec{x}$, which are called explanatory variables. More formally, the learning problem is the calculation of a vector $\beta$ such that
\begin{equation*}
    y = \vec{x}^T\beta + \epsilon.
\end{equation*} 

Least-squares method is a popular approach for estimating $\beta$. Let $X$ be the design matrix with $n$ data samples and $Y$ be the corresponding vector of labels. The model parameters are then given by
\begin{equation}
\label{eq:beta}
    \beta = (X^TX)^{-1} X^TY.
\end{equation} 

Let $X_i$ and $Y_i$ be a horizontal partitioning of the design matrix and label vector, respectively. Each user $i$ only has access to $X_i$ and $Y_i$. As noted by the authors in~\cite{fang2013privacy}, the following equations hold

\begin{gather*}
    X^TX = \sum_i X_i^TX_i,\\
    X^TY = \sum_i X_i^TY_i. 
\end{gather*}
Therefore,
\begin{equation}
   \beta = (\sum_i X_i^TX_i)^{-1} \sum_i X_i^TY_i.
\end{equation}

\noindent
\textbf{Computing $\mathbf{\beta}$:} Let the dimension of the data $(\vec{x})$ be $d$. Each user independently computes a $d\times d$ matrix $(X_i^TX_i)^{-1}$ and a $d$ dimensional vector $X_i^TY_i$. The encrypted tensors are submitted along with range proofs on the $L^2$-norm to bound the influence of each user on the final model parameters. The tensors are homomorphically added to calculate $\beta$ as per equation~\ref{eq:beta}.

\subsubsection{Collaborative Filtering}
\label{apps:cf}
Collaborative Filtering (CF) is a technique most commonly used in recommender systems to predict the preferences of a user by accumulating preferences of multiple users. Among the various approaches that exist in literature for CF~\cite{rbm, goldberg2001eigentaste,canny2002collaborative}, we focus on the one used by Canny~\cite{canny2002collaborative}. This work uses the ideas of secret sharing and threshold decryption to achieve CF with privacy. It relies on a majority vote among untrusted tallying authorities to get the result of the computation. In contrast, our approach gets rid of the tallying authorities by carefully designing the encryptions. We only mention the key computation steps of the algorithm by Canny; interested readers should refer to~\cite{canny2002collaborative} for details.

Let there be $n$ users providing integer ratings to $m$ items. Let $P^{n\times m}$ be the user preference matrix such that $P_{ij}$ is the rating given by user $i$ to item $j$. $P_{ij}$ is $0$ if the item is unrated. The first step is the derivation of a low dimensional approximation to $P$. Let $A^{k\times m}$ ($k$ is small) be such an approximation:
\begin{equation*}
    A = \sup_{U: UU^T=I} \tr(PU^TUP^T).
\end{equation*}
Starting from a random matrix, $A$ is computed iteratively using conjugate gradient. Let $A_{(t)}$ be value of the matrix at iteration $t$, and $P_i$ denote the $1\times m$ matrix of data from user $i$. The gradient for the current iteration can be calculated as 
\begin{equation*}
    G_{(t)} = \sum_{i=1}^{n} A_{(t)}{P_i}^TP_i(I-A_{(t)}^TA_{(t)}).
\end{equation*}
After $A_{(t)}$ is updated using the gradient, the process is repeated (until convergence). Generating new recommendations from $A$ entails more steps like partial SVD and probabilistic latent variable modeling~\cite{canny2002collaborative}. \\

\noindent
\textbf{Computing $\mathbf{G}$ in every iteration:} Since $G = \sum_i G_i$, we can use homomorphic encryption to securely calculate the gradient in a distributed setting. Each user creates an encrypted matrix ${G_i}^{k\times m}$. To limit the effect of each $G_i$ on the final gradient, a range proof on the $L^2$-norm of $G_i$ is required. 

\clearpage

\subsection{Evaluations}
\label{appx:eval}

\paragraph{ElGamal Encryption and \zkp ~libraries:} To achieve higher efficiency, we write our own lightweight \zkp ~libraries instead of using existing general \zkp ~libraries such as zk-SNARK~\cite{libsnark}. We implement ElGamal encryption over elliptic curve using Jeremy Kun's elliptic curve library~\cite{Kun2014} in Python. For each proof mentioned in Appendix~\ref{appx:zkp}, we implement a \zkp ~library to generate and verify the proof based on ElGamal encryption. 

\paragraph{Zorro Client:} This is an interfacing client that takes an input vector $(T_i)$ from the application, and returns a vector summation $(\sum_i T_i)$ computed over all the parties involved in the protocol. Developers who want to implement applications in Table~\ref{table:appssummary} can use the Zorro client as a black-box and do not need to be aware of the underlying \zkp s or the interactions with the public ledger. Specifically, the Zorro client handles the following 3 tasks for the higher level application:
\begin{itemize}
    \item Generate \zkp s for input validity and commit them to the ledger;
    \item Verify \zkp s commited by other users;
    \item Calculate vector summation by homomorphic vector addition over encrypted inputs of all the users.
\end{itemize}

To evaluate the efficiency of Zorro client, we measure the \zkp ~generation and verification time for each user for the application of cumulative voting. Considering that users of the application (i.e., voters) would not have access to specialized hardware, the evaluations are done on a regular laptop with 2.7 GHz Intel Core i5. The Zorro client implements the two-round homomorphic vector addition protocol introduced in Section~\ref{subsec:two-round-pro}. The first round of the protocol consists of one \zkp ~of discrete log (Eq.~\ref{zkpeq1}) for each element in the input vector. The time complexities of \zkp ~generations and verifications for the first round increase linearly with the length of vector, and do not depend on any $L^1$-norm constrains on the input. Therefore, we focus our evaluations on the second round of the protocol. 

\begin{figure*}[h!]
    \begin{minipage}[b]{.5\linewidth}
        \centering
        \includegraphics[width=\columnwidth]{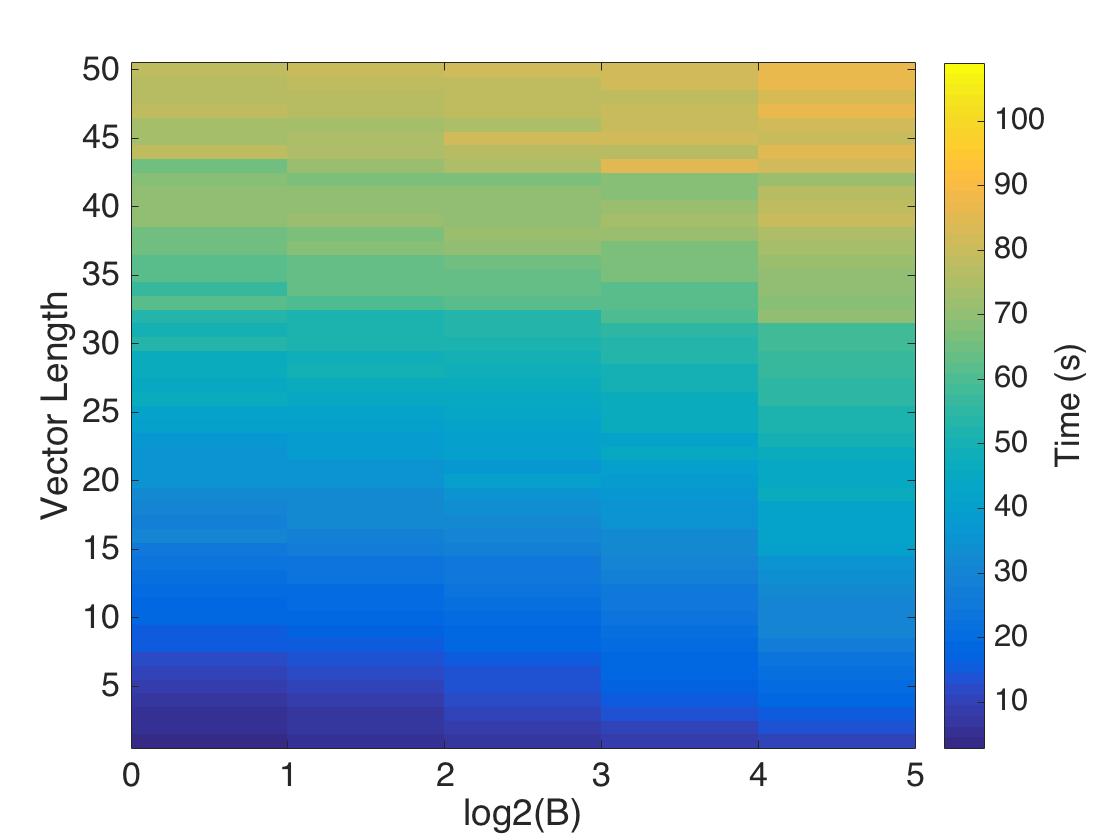}%
        \subcaption{\zkp ~generation time per user.}
        \label{fig:prftime}
    \end{minipage}%
    \begin{minipage}[b]{.5\linewidth}
        \centering
        \includegraphics[width=\columnwidth]{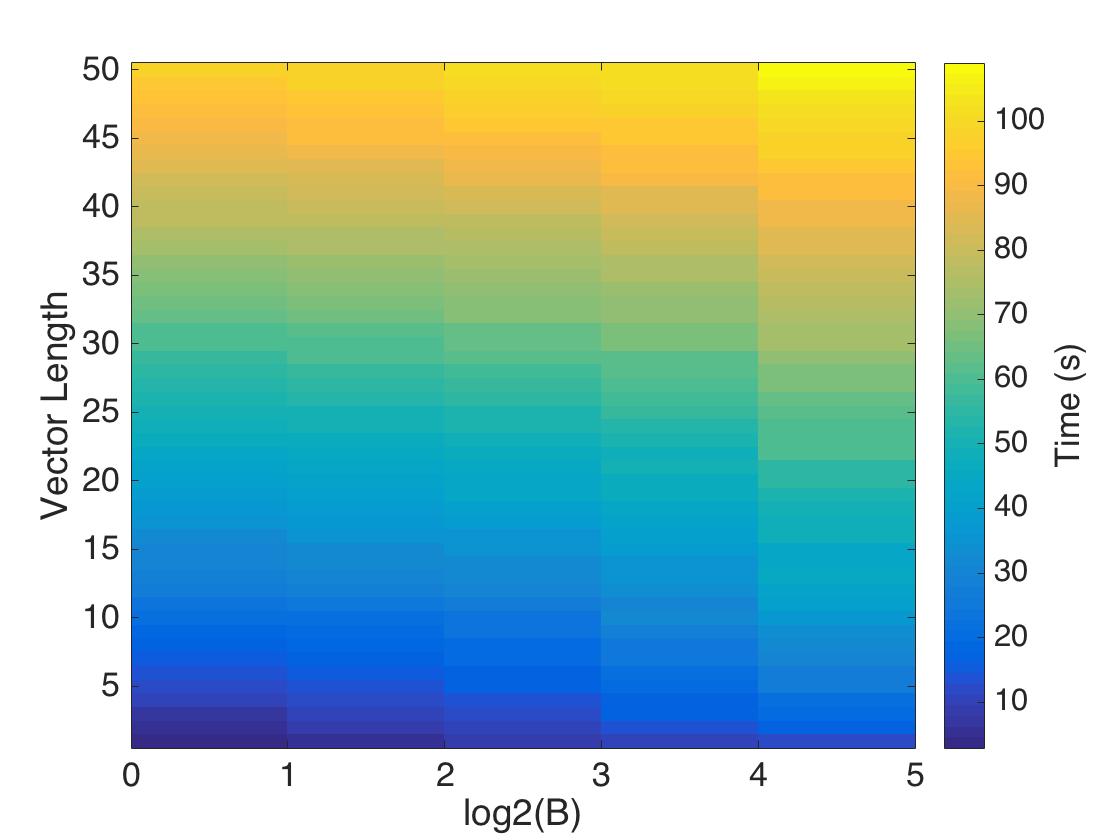}%
        \hfill
        \subcaption{\zkp ~verification time per user. ($n=1$)}
        \label{fig:vrftime}
    \end{minipage}
\caption{Time Complexity Analysis under Varying Vector Length and Maximum Bound}
\label{fig:time}
\end{figure*}

\paragraph{\zkp ~Generation Time} The \zkp ~generation time in round 2 depends on two factors: vector length ($m$) and maximum bound ($B$) on the $L^1$-norm. For cumulative voting, the vector length corresponds to the number of candidates, and the maximum bound corresponds to the number of votes per voter. The former determines the number of range proofs one client needs to generate, while the latter determines the complexity of each range proof. Figure~\ref{fig:prftime} shows the variation of \zkp ~generation time per client for vector length $1 \le m \le 50$ and maximum bound $2^1 \le B \le 2^5$. The figure reveals positive correlations between \zkp ~generation time and vector length, and between \zkp ~generation time and maximum bound. Out of vector length and maximum bound, we observe the impact of the former to be higher. For example, when $m=1$, $B=2^5=32$, it takes only $9.9$ seconds to generate \zkp s. However, when $m=32$, $B=2^1$, the generation time takes around $53.2$ seconds. Therefore, Zorro can handle cumulative voting with relatively large number of votes per user, but is more suitable for a small number of candidates. When the number of candidates exceeds 35, it takes more than one minute to generate the \zkp s even when only one vote is allowed per voter. 

\begin{figure*}[h!]
\centering
\includegraphics[width = 0.5\columnwidth]{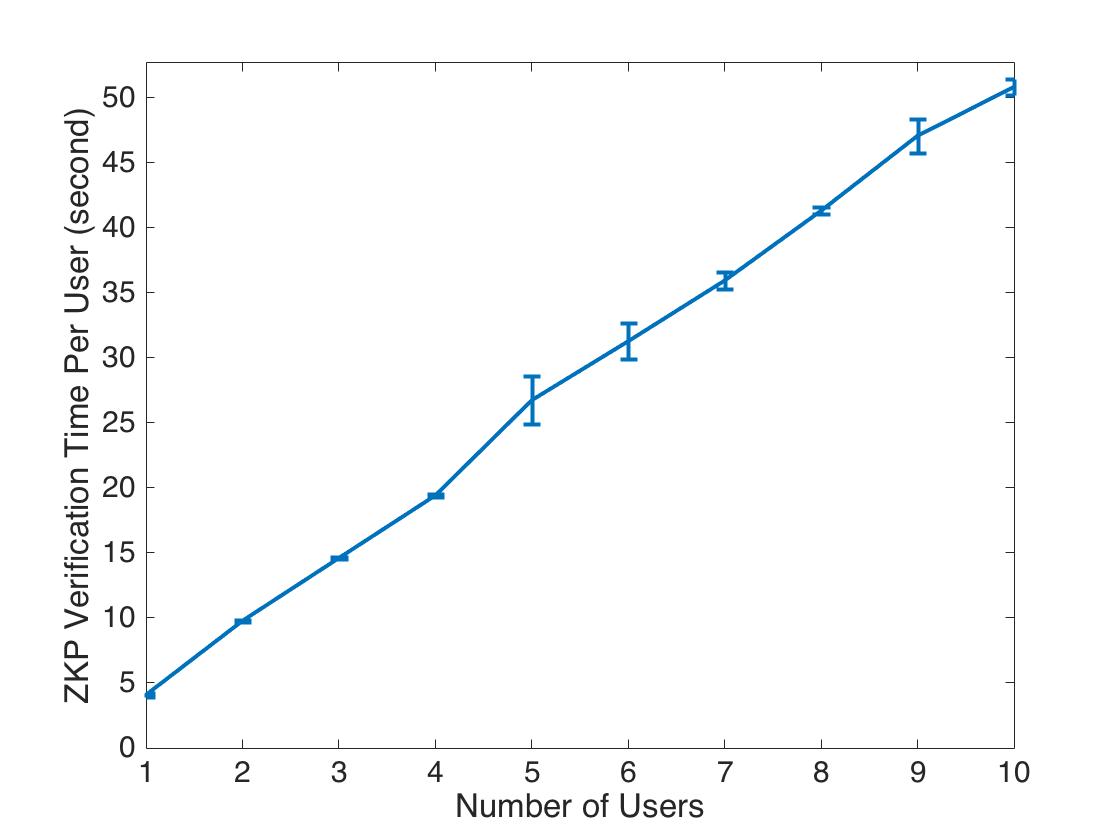}%
\hfill
\caption{\zkp ~Verification Time Per User against Increasing Number of Users ($m = 1$, $B = 2$)}
\label{fig:time_usr}
\end{figure*}

\paragraph{\zkp ~Verification Time} The \zkp ~verification time depends on three factors: vector length ($m$), maximum bound ($B$), and total number of users ($n$). Since each client needs to verify \zkp s of \emph{all} the users, \zkp ~verification time per client increases linearly with the number of users. Figure~\ref{fig:time_usr} shows the increase in average per-client verification time (when $m=1$ and $B=2$) as the number of total users increases from 1 to 10. On average, it takes around 5 seconds to verify the \zkp s of each user. Therefore, when there are thousands of users, the verification phase can take hours. However, since the verification for different users is independent, the overall time can be greatly reduced by using multi-core parallelism. Furthermore, the optimization techniques discussed in section~\ref{subsubsec:opt} can also be applied to improve efficiency. Similar to \zkp ~generation time, \zkp ~verification time is also influenced by vector length and maximum bound. Figure~\ref{fig:vrftime} shows \zkp ~verification time with $1 \le m \le 50$, $2^1 \le B \le 2^5$, and $n=1$. On average, the time it takes to verify \zkp s is slightly higher than the time it takes to generate them. 

\begin{figure*}[h!]
    \begin{minipage}[b]{.5\linewidth}
        \centering
        \includegraphics[width=\columnwidth]{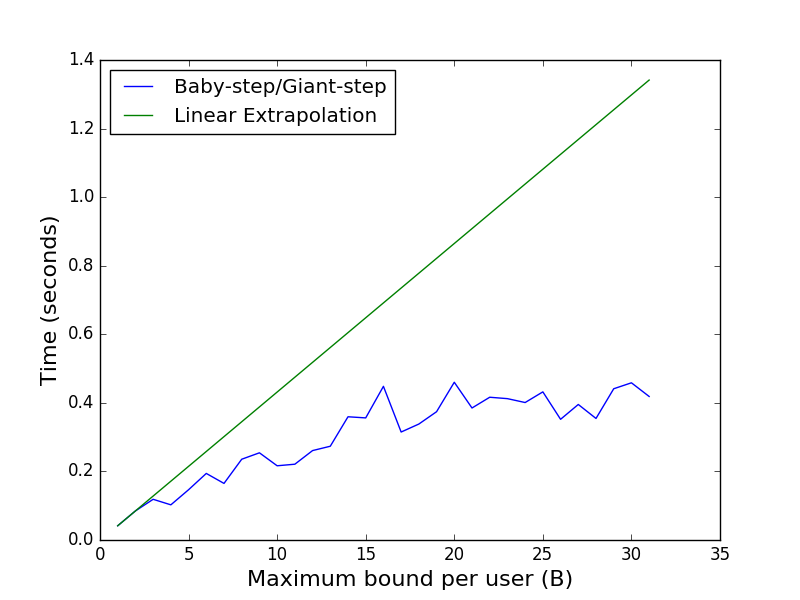}%
        \subcaption{Time to compute discrete log using baby-step/giant-step}
        \label{fig:time_bg}
    \end{minipage}%
    \begin{minipage}[b]{.5\linewidth}
        \centering
        \includegraphics[width=\columnwidth]{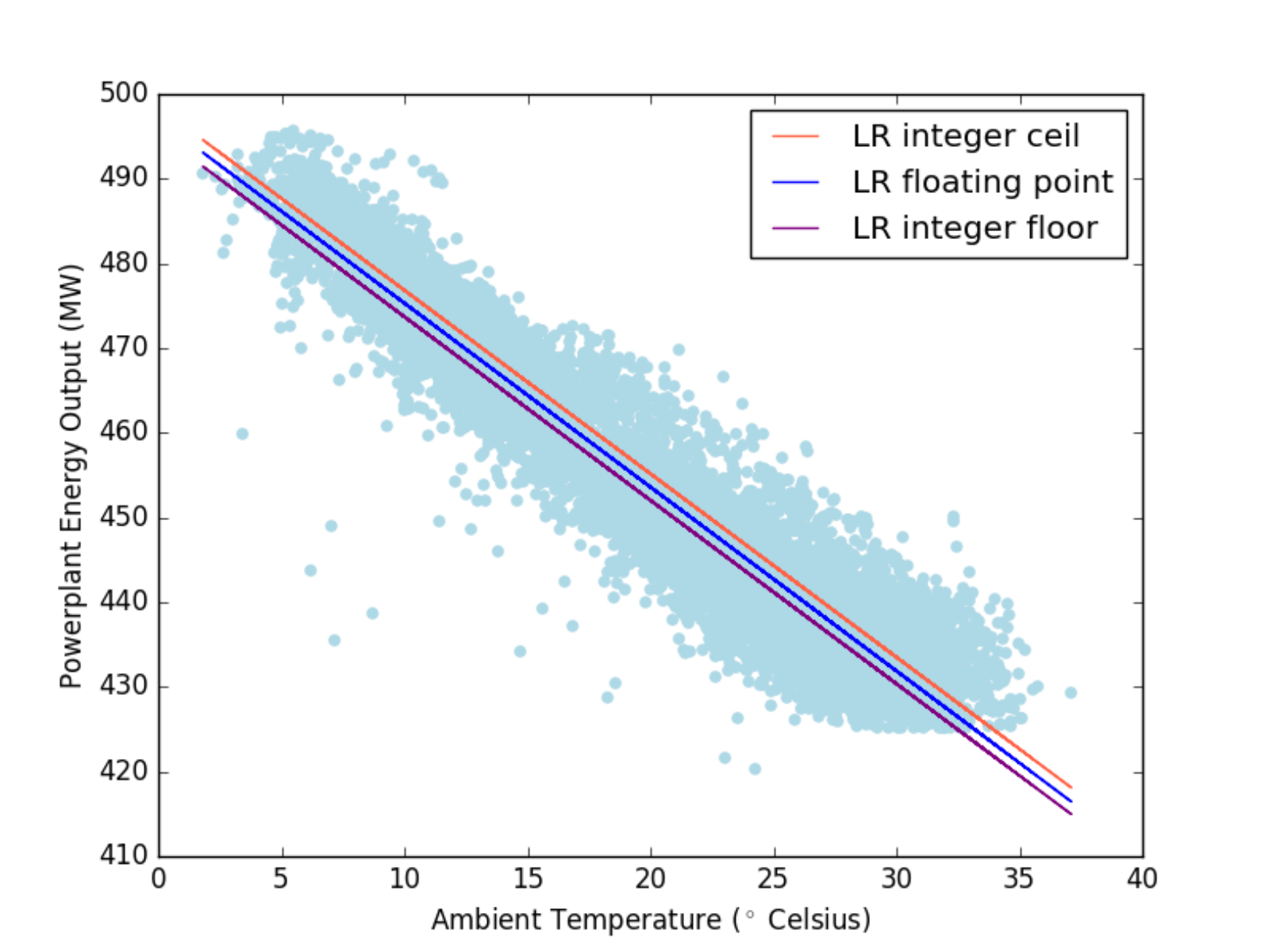}%
        \hfill
        \subcaption{Power output prediction using linear regression}
        \label{fig:linearReg}
    \end{minipage}
\caption{(a) Analysis on discrete log computation and (b) Accuracy loss with linear regression}
\label{fig:time}
\end{figure*}

\paragraph{Taking the Discrete Log} As mentioned in Section~\ref{subsec:two-round-pro}, each user calculates $g^{\sum_i T_{ij}}$ by multiplying the 
correct ciphertexts, and uses the baby-step/giant-step algorithm~\cite{lenstraalgorithms} to obtain the
discrete log. The algorithm has a time complexity of $O(\sqrt{N})$, for a search space of 
$N$ numbers. In Figure~\ref{fig:time_bg}, we plot the time to compute the discrete log as a function 
of the bound on the input from each user. We simulate 1000 users, each with an integer input $T_{ij}$ in the 
range $[0, B]$, generated using a uniform distribution. It then follows that the sum $\sum_i T_{ij}$ is a
value in the range $[0,1000\times B]$, distributed according to a Irwin-Hall distribution. We record the 
time taken to compute the discrete log of the sum, average it over 10 observations and plot. Figure~\ref{fig:time_bg} shows that the algorithm has sub-linear time complexity. Moreover, the discrete log can be calculated in less than a second even with $B$=32. Hence, this step is very fast compared to the \zkp\ generation and verification steps mentioned above.

\paragraph{Impact on Accuracy} Our elliptic curve cryptography system uses a finite field of integers modulo p, $\mathbb{Z}_p$. Therefore, the input vectors to our homomorphic vector addition algorithms can only be integers from this field. Although sufficient for cumulative voting, this may be restrictive for some machine learning applications which are sensitive to floating point (FP) precision. We evaluate uni-variate linear regression (Section~\ref{apps:lr}) on a real data-set and quantify the loss. Figure~\ref{fig:linearReg} plots the variation of the electrical power output from a power plant with ambient temperature~\cite{tufekci2014prediction}. The input and output values have FP precision. We fit a linear regression model to the data in three ways, first by using the original values, and then by using \texttt{floor} and \texttt{ceil} on the FP data in two separate experiments. We observe that \texttt{floor} and \texttt{ceil} models have 8.3\% and 8.4\% higher mean square error than the FP model, respectively. This shows that the loss in accuracy due to FP rounding-off errors can be small. Furthermore, we can use FP quantization methods to improve precision, if needed.

\end{document}